# Simple and accurate exchange energy for density functional theory

Teepanis Chachiyo,[1,2,*] and Hathaithip Chachiyo[3]

[1] Department of Physics, Faculty of Science, Naresuan University, Phitsanulok 65000 Thailand
email: teepanisc@nu.ac.th

[2] Thailand Center of Excellence in Physics, Ministry of Higher Education, Science, Research and Innovation, 328 Si Ayutthaya Road, Bangkok 10400, Thailand

[3] Private Residence, Phitsanulok 65000, Thailand email: hathaithip.chachiyo@gmail.com

[*] Corresponding author

## Abstract

A non-empirical exchange functional based on an interpolation between two limits of electron density: slowly varying limit and asymptotic limit, is proposed. In the slowly varying limit, we follow the study by Kleinman in 1984 which considered the response of a free-electron gas to an external periodic potential, but further assume that the perturbing potential also induces Bragg diffraction of the Fermi electrons. The interpolation function is motivated by the exact exchange functional of a hydrogen atom. Combined with our recently proposed correlation functional, tests on 56 small molecules show that, for the first-row molecules, the exchange-correlation combo predicts the total energies four times more accurate than presently available Quantum Monte Carlo results. For the second-row molecules, errors of the core electrons exchange energies can be corrected, leading to the most accurate molecular total energy predictions to date despite minimal computational efforts. The calculated bond energies, zero point energies, and dipole moments are also presented.

# I. INTRODUCTION

Exchange energy plays an important role in density functional theory (DFT).[1] Even though much has been understood regarding its behavior,[2,3] there is plenty of room for more understanding. For example, it is well-established that the exchange energy obeys two limiting cases: (i) the slowly varying density, and (ii) the asymptotic limit where the density decays rapidly in the outer regions of molecules.[4] In 1988, Becke proposed a functional that satisfied the two limits.[2] The question remains, however, what exactly happens in the intermediate region? In this work, we explicitly use an interpolation function to merge the two limits, and show that the resulting functional is simple and accurate.

Another example is related to the slowly varying limit where the exchange energy increases as a function of the reduced gradient parameter $s \equiv |\vec{\nabla}\rho|/2(3\pi^2)^{1/3}\rho^{4/3}$.[3] Although it is well known that the exchange energy enhancement factor increases quadratically $(1 + \mu s^2)$, the value of $\mu$ is not exactly understood.[5] An often cited value comes from a study by Klienman which considered how free-electron eigenfunctions were perturbed by an external periodic potential.[6] The exchange energy was then evaluated and rewritten as a function of electron density and its gradient. If the wavelength of the external potential was large, it was estimated that $\mu = \frac{8}{81}$ which was too low to give accurate predictions for atoms and molecules.[5] In this work, we assume an additional condition on the external periodic potential. That is, the potential also induces Bragg diffraction of the Fermi electrons. This assumption leads to a higher value of $\mu$ and gives accurate predictions of total energies, bond energies, zero point energies, and dipole moments of test systems, which include first- and second-row atoms and 56 small molecules.

This report is organized as follow. In Theoretical Method, we layout the mathematical form, propose the interpolating function, derive the value of $\mu$, and fully construct the exchange functional. Then, we describe benchmarking procedures. In Results and Discussions, we show that the exchange functional is accurate, particularly in comparison to the previously published Quantum Monte Carlo (QMC) calculations.[7] In Conclusion, we list all the theoretical arguments that lead to the non-empirical exchange functional in this work.

## II. THEORETICAL METHOD

Traditionally, a dimensionless parameter $s$ was used to quantify the inhomogeneity of electron density. The exchange energy was then written with an "enhancement factor" $F(s)$,[3]

$$E_x[\rho] = \int \rho \varepsilon_x F(s) d^3r. \quad (1)$$

Here, $\varepsilon_x = -\frac{3}{4}\left(\frac{3}{\pi}\rho\right)^{1/3}$ was the Dirac exchange energy density[8] for uniform electron gas. As the electron density deviated from the homogeneous case $s = 0$, the function $F(s)$ increased from the baseline value of 1, effectively enhancing the strength of the exchange energy from that of the uniform electron gas.

As March elaborated,[9] in the asymptotic limit $s \to \infty$, the exchange energy density needed to behave as $\varepsilon_x F(s) \to -1/2r$. Also asymptotically, the electron density decayed exponentially $\rho(r) \to Ne^{-ar}$. This asymptotic behavior was not specific to a hydrogen atom but occurred in general,[4] where the parameter $a$ was related to the ionization potential. Therefore, to satisfy the limit, the enhancement factor was thought to take the form $F(s) = \frac{cs}{\ln s}$, which however, diverged to infinity at $s = 1$. Therefore, as the first step, we proposed the following modification.

$$s \to \infty: \qquad F(s) = \frac{cs}{\ln(cs+1)}; \quad c = \frac{4\pi}{9}, \quad (2)$$

which was well behaved in the entire range $s \in [0, \infty)$ with an additional advantage: $F(0) = 1$. It was easy to show that for the $\varepsilon_x F(s) \to -1/2r$ in this limit, the constant $c$ had to be equal to $4\pi/9$.

The next step was to merge $F(s)$ in Eq. (2) with the slowly varying case. It was known[5,6] that in this region the enhancement factor grew quadratically as $1 + \mu s^2$. Ignoring altogether the $\mu s^2$ behavior in the beginning and conjecturing that the quadratic dependence would have emerged naturally after the interpolation, we simply tried to interpolate between the *perfectly uniform* electron gas $F(s) = 1$ in the $(s = 0)$ limit, and the $F(s) = \frac{cs}{\ln(cs+1)}$ in the asymptotic limit $(s \to \infty)$ with the simplest interpolation scheme possible, namely

$$F(s) = 1 \cdot w(s) + \frac{cs}{\ln(cs+1)} \cdot [1 - w(s)]. \tag{3}$$

Here, $w(s)$ was the weighting function which took a value between $[0,1]$. We used $w(s) = \frac{1}{ds+1}$ with the constant $d$ controlling how rapidly the weighting function migrated from the slowly varying limit to the asymptotic limit. Substituting the $w(s)$ into Eq. (3) yielded

$$F(s) = \frac{dcs^2 + \ln(cs+1)}{(ds+1)\ln(cs+1)}. \tag{4}$$

The choice of the weighting function $w(s) = \frac{1}{ds+1}$ was motivated by the exact exchange functional for a hydrogen atom. Consider a hydrogen atom whose normalized electron density was $\rho(r) = \frac{1}{\pi} e^{-2r}$. Recall that the electron repulsion energy $E_{e-e} = \frac{1}{2} \iint d^3 r_1 d^3 r_2 \frac{\rho(\vec{r}_1)\rho(\vec{r}_2)}{|\vec{r}_1 - \vec{r}_2|}$ of the Kohn-Sham DFT was always there in the formalism regardless of the physical system under study. However, there was only one electron in hydrogen atom; hence, the role of the exchange energy in this case was to cancel exactly the $E_{e-e}$ in the Kohn-Sham formalism. For a hydrogen atom,

$$E_{e-e} = \int d^3 r \rho(\vec{r}) \left[ e^{-2r}(-\tfrac{1}{2}) + (1 - e^{-2r}) \frac{1}{2r} \right] \ . \tag{5}$$

Therefore, the exchange energy density which cancelled the above expression exactly was

$$\varepsilon_x(\rho)F(s) = e^{-2r}(\tfrac{1}{2}) + (1 - e^{-2r})\left(-\tfrac{1}{2r}\right). \tag{6}$$

Notice the curious pattern which mimicked an interpolation with the term $e^{-2r}$ representing the weighting function. Note that weighting function decayed exponentially in $r$.

Using $\rho(r) = \frac{1}{\pi}e^{-2r}$, we had $s \propto e^{2r/3}$. In other words, $\frac{1}{ds+1}$ also decayed exponentially in $r$, approximately mimicking the behavior of the exact exchange functional of a hydrogen atom. In fact, even if we had allowed a more general form $w(s) = \frac{1}{ds^n+1}$, the second order Taylor expansion of $F(s)$ in Eq. (3) would have vanished for all positive integers $n$, except when $n = 1$. In other words, $n = 1$ was the only case consistent with the quadratic behavior of the enhancement factor. Therefore, we believed $w(s) = \frac{1}{ds+1}$ was the optimal form of the weighting function.

Next, we determined the value of $\mu$ in the slowly varying limit. In generalize gradient approximation, when the gradient parameter $s$ was small, the exchange energy (per unit volume) was approximately

$$\frac{E_x}{V} = -\frac{3}{4}\left(\frac{3}{\pi}\right)^{1/3}\rho^{4/3}(1 + \mu s^2). \tag{7}$$

In 1984, Kleinman studied how free-electron eigenfunctions were perturbed by an external periodic potential $V_{\text{ext}}(\vec{K})e^{+i\vec{K}\cdot\vec{r}} + V_{\text{ext}}(-\vec{K})e^{-i\vec{K}\cdot\vec{r}}$.[6] To avoid confusion, we will use the notation Eq. (K-Number) when referring to a specific equation in the Kleinman's paper. From Eq. (K-17), the exchange energy per normalization volume Ω was

$$\frac{E_x}{\Omega} = -\frac{3}{4}\left(\frac{3}{\pi}\right)^{1/3}\rho_0^{4/3} - \frac{2\pi}{3K^2}\left\{\left[\rho(\vec{K})\right]^2 + \left[\rho(-\vec{K})\right]^2\right\}. \tag{8}$$

Using $\rho(\vec{K}) = \rho(-\vec{K})$ as indicated in the Eq. (K-10), we had

$$\frac{E_x}{\Omega} = -\frac{3}{4}\left(\frac{3}{\pi}\right)^{1/3} \rho_0^{4/3} \left(1 + \frac{4}{3}\left(\frac{\pi}{3}\right)^{1/3} \rho_0^{-4/3} \frac{2\pi}{3K^2} 2\left[\rho(\vec{K})\right]^2\right). \tag{9}$$

Comparing Eq. (7) and Eq. (9), we could identify

$$\frac{4}{3}\left(\frac{\pi}{3}\right)^{1/3} \rho_0^{-4/3} \frac{2\pi}{3K^2} 2\left[\rho(\vec{K})\right]^2 = \mu s^2 = \mu \frac{1}{4(3\pi^2)^{2/3}} \frac{\left|\vec{\nabla}\rho\right|^2}{\rho^{8/3}}. \tag{10}$$

Using $\rho \approx \rho_0$ as generally the case for perturbed free-electron gas, and $k_F = (3\pi^2\rho_0)^{1/3}$, we had

$$\mu = \frac{32}{27} \frac{k_F^4}{K^4} \frac{2K^2\left[\rho(\vec{K})\right]^2}{\left|\vec{\nabla}\rho\right|^2}. \tag{11}$$

It was possible to mathematically deduce that $\frac{2K^2\left[\rho(\vec{K})\right]^2}{\left|\vec{\nabla}\rho\right|^2} = 1$ on average (see supplementary material). Therefore,

$$\mu = \frac{32}{27}\left(\frac{k_F}{K}\right)^4. \tag{12}$$

One choice of $\left(\frac{k_F}{K}\right)^4$ was of particular interest. Consider the case when the external potential induced Bragg diffraction of the Fermi electrons. The mathematical condition for this to occur was[10]

$$2\vec{k}_F \cdot \vec{K} = K^2 \quad \text{or} \quad 4k_F^2 K^2 \cos^2(\theta) = K^4. \tag{13}$$

Replacing $\cos^2(\theta)$ with its average value $\frac{1}{\pi}\int_0^{\pi}\cos^2(\theta)d\theta = \frac{1}{2}$, and rearranging the above expression, we had $\left(\frac{k_F}{K}\right)^4 = \frac{1}{4}$. Hence,

$$\mu = \frac{8}{27} = \frac{24}{81}. \tag{14}$$

Having elaborated on the choice of the weighting function, and determined the value of $\mu$, we revisited Eq. (4) and attempted to compute the constant $d$ using Taylor expansion up to the second order:

$$s \ll 1: \qquad F(s) \approx 1 + 0 \cdot s + \frac{1}{2}(dc)s^2, \tag{15}$$

which could readily be compared to the known $1 + \mu s^2$ behavior of the slowly varying limit. We used the value $\mu = \frac{8}{27}$ from Eq. (14), yielding

$$\frac{1}{2}(dc) = \frac{8}{27} \quad \text{or} \quad d = \frac{8}{27}\frac{2}{c} = \frac{4}{3\pi}. \tag{16}$$

Putting the constant $d$ and $c$ back into Eq. (4) and defining another variable $x \equiv cs = \frac{|\nabla\rho|}{\rho^{4/3}}\frac{2}{9}\left(\frac{\pi}{3}\right)^{1/3}$, we finally arrived at the non-empirical exchange functional in this work

$$E_x = \int \rho\varepsilon_x \frac{3x^2+\pi^2\ln(x+1)}{(3x+\pi^2)\ln(x+1)} d^3r. \tag{17}$$

To produce the results as shown in the following section, one also needed a correlation energy contribution. We used our recently published correlation functional[11]

$$E_c = \int \rho \varepsilon_c (1 + t^2)^{h/\varepsilon_c} \, d^3 r. \tag{18}$$

This exchange-correlation combo was shown to give very accurate total energies for a small set of molecules. Here, we expanded the size of test molecules, and the variety of atoms to cover second-row elements.

To allow direct comparisons with the previous QMC results,[7] we used a set of 56 molecules with the exact same geometries as described in the supplementary material of Ref. 7. Experimentally derived total energies[12] and zero point energies[17] were also available for comparisons. We used a large basis set QZP[13]-gh (with g and h GTO removed) for single point calculations, and a smaller basis 6-31G* for zero point energy calculations. The basis sets were downloaded from EMSL Basis Set Exchange database.[14] The reported values were calculated using the Siam Quantum program.[15] The detailed methods for solving the Khon-Sham SCF equation for this exchange-correlation functionals were described in the supplementary material of Ref. 11. Very recently, the functionals in Eq. (17) and Eq. (18) were implemented in the LibXC library[16] where the calculated energies agreed with that of the Siam Quantum's implementation to within micro-hartrees.

## III. RESULTS AND DISCUSSIONS

Fig. 1 illustrates the accuracy of the exchange functional in this work when predicting the total energies of neutral atoms. The label "(This Work) DFT" is from the exchange and correlation functional in Eq. (17) and Eq. (18), respectively. The performance is satisfactory for small atoms, but increasingly poorer starting from Ne onward. The results from QMC calculation[7] also show the same trend.

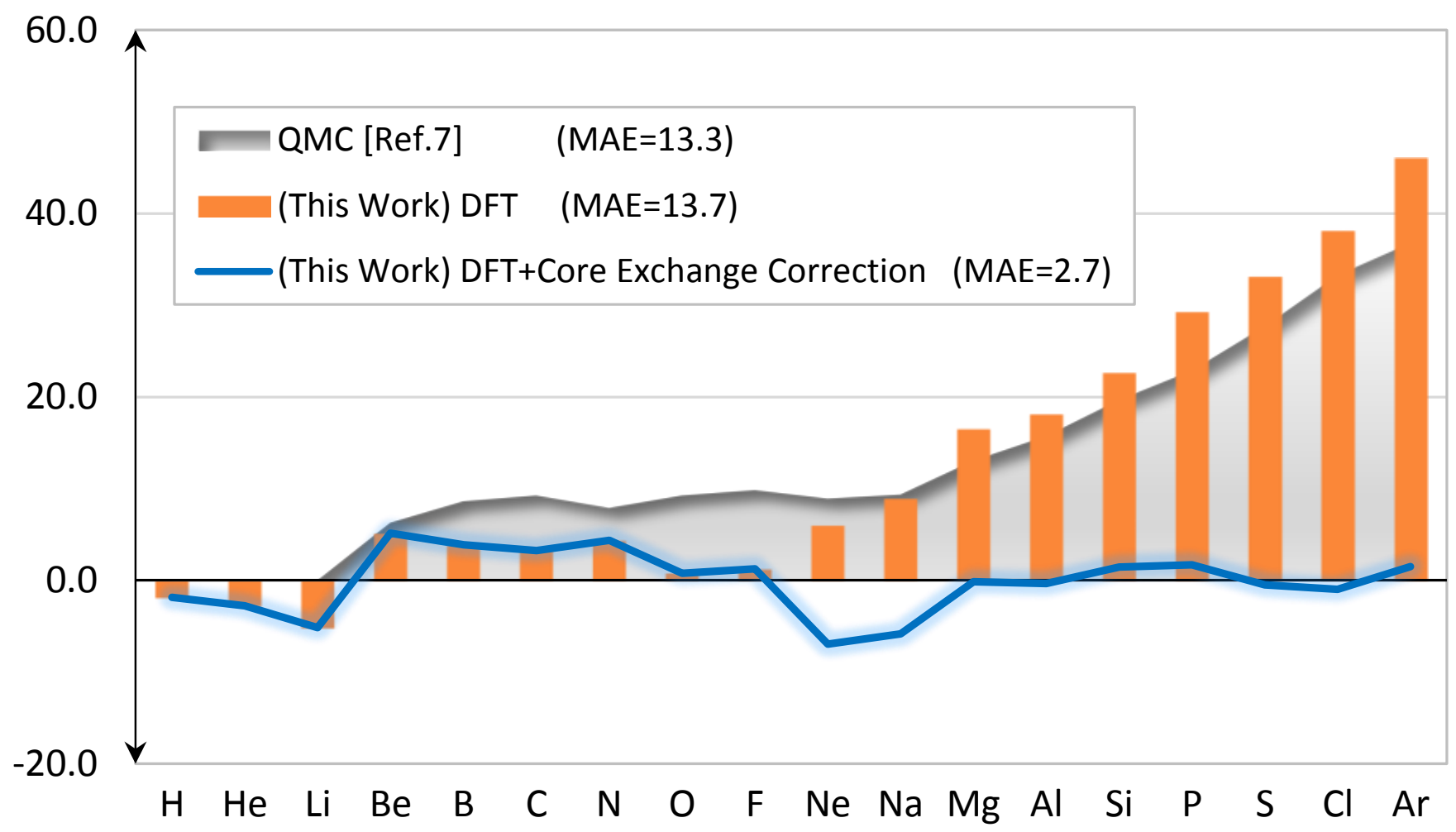

FIG. 1. Errors of the calculated total energies as compared to the "exact" values[12] (kcal/mol). The QMC results[7] are also shown.

After a few trial-and-error attempts, we discover that the errors of the predicted total energies for Ne and the second-row atoms are related to the core electrons' exchange energy. A simple ad-hoc correction scheme for atoms and molecules can be devised very accurately. As shown in Fig. 2, we first compute the error of exchange energy at Hartree-Fock (HF) density for each ion core. For example, HF calculation is performed on $Ne^{8+}$, yielding HF orbitals and HF density. Then, the orbitals are used to compute HF exchange energy; whereas the density is used to compute the DFT exchange energy via Eq. (17). The difference $(\mathrm{DFT}_x - \mathrm{HF}_x)$ is reported in Fig. 2.

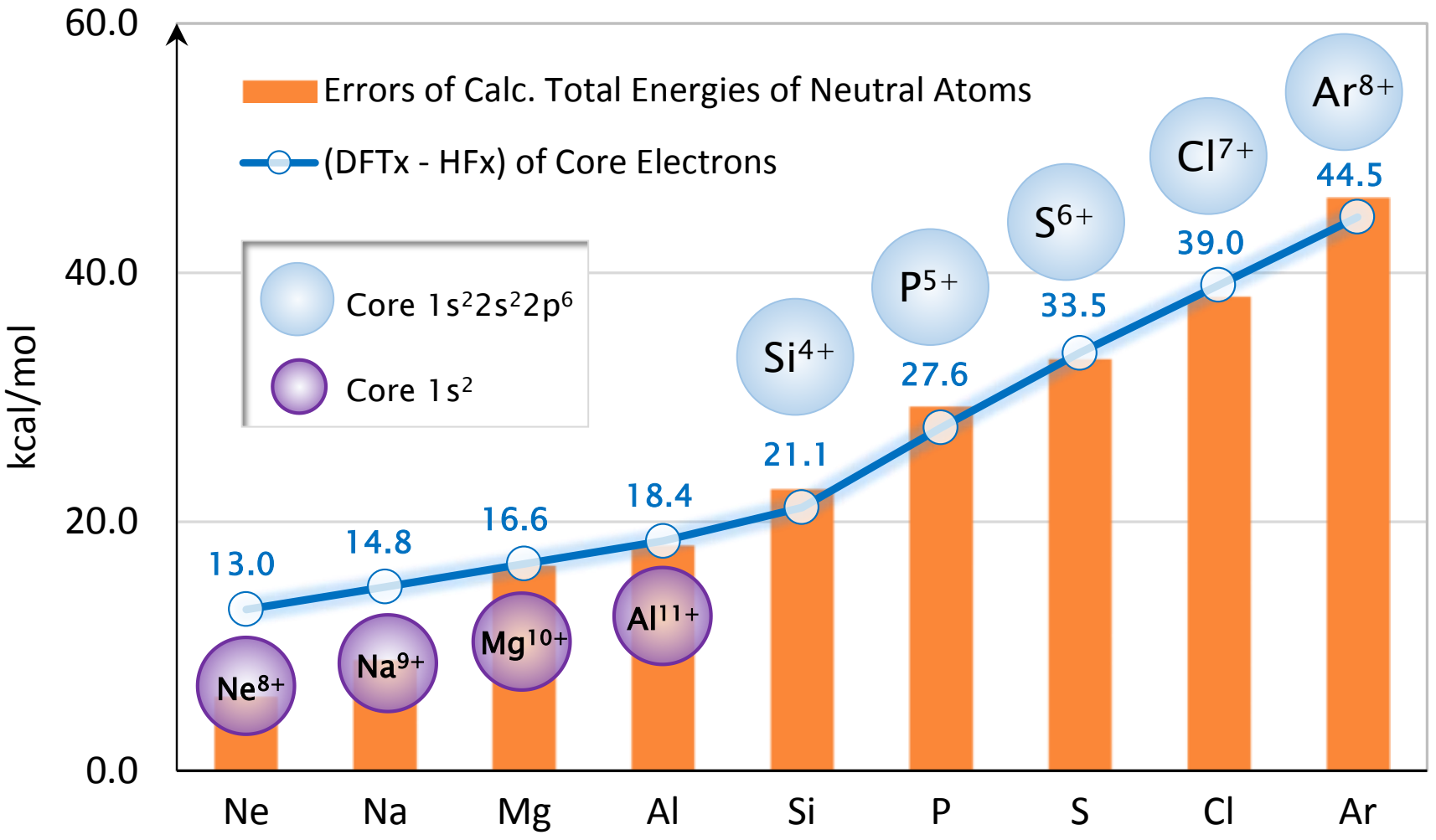

FIG. 2. Schematic of the Core Exchange Correction.

From the graph, $(\mathrm{DFT}_x - \mathrm{HF}_x)$ of the core electrons track quite accurately with the errors of the predicted total energies. Therefore, we devise an error correction scheme by simply subtracting the $(\mathrm{DFT}_x - \mathrm{HF}_x)$ of the core electrons from the DFT's total energy. In other words,

$$\text{Core Exchange Correction } \Delta E_{\mathrm{x}}^{(\mathrm{core})} \equiv -\left(\mathrm{DFT}_x^{(\mathrm{core})} - \mathrm{HF}_x^{(\mathrm{core})}\right). \tag{19}$$

The corrected total energies, $E_{\mathrm{total}}^{(\mathrm{DFT})} + \Delta E_{\mathrm{x}}^{(\mathrm{core})}$, are shown in Fig. 1. The mean-absolute-error (MAE) is now dropped to only 2.7 kcal/mol. The list of $\left(\mathrm{DFT}_x^{(\mathrm{core})} - \mathrm{HF}_x^{(\mathrm{core})}\right)$ is given in Fig. 2. In this ad-hoc scheme, the correction is only applied to Ne through Ar.

The main results of this report are the errors of the calculated total energies of 56 molecules as shown in Fig. 3. Even without any correction, the DFT energies perform very well for the first row molecules with MAE of 3.2 kcal/mol, four times more accurate than the presently available QMC results. However, when the test molecules contain second-row atoms, the errors rise to match that of the QMC. After we apply the correction scheme, the errors suddenly drop; and the MAE of the full set is now only 3.5 kcal/mol.

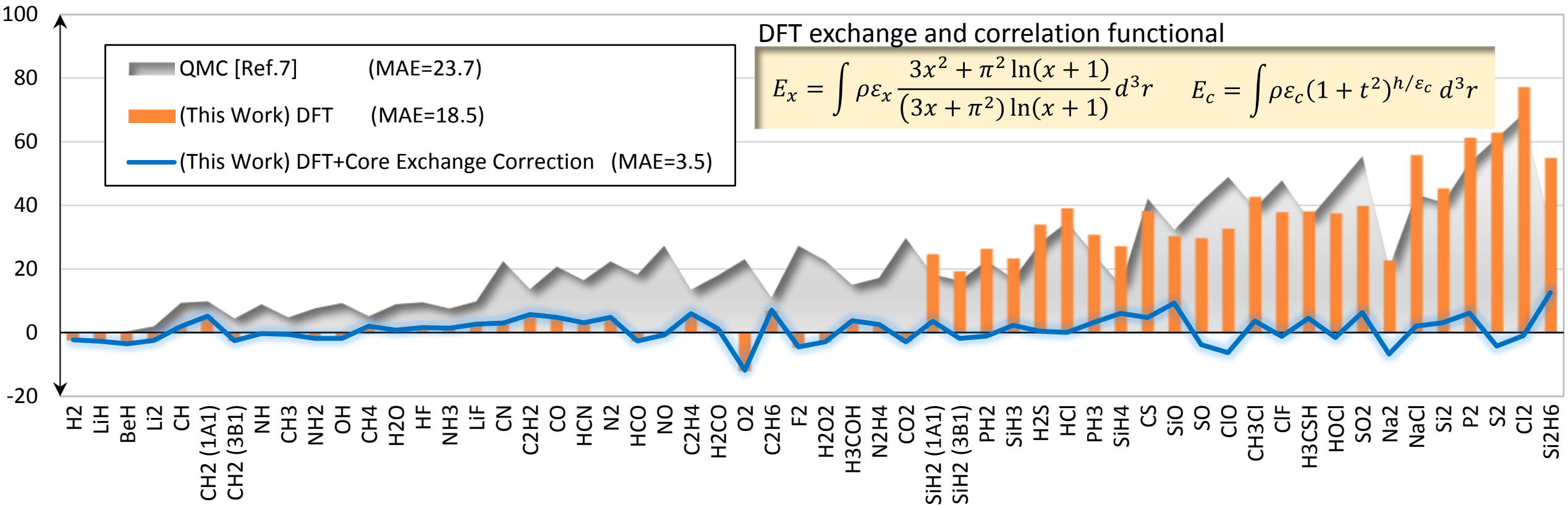

FIG. 3. Errors of calculated total energies (kcal/mol). The reference total energies[12] are derived from experimental atomization energies.

For molecules, the correction is done “atom-wise”, and only on the second-row atoms. That is,

$$\text{Corrected Molecular Total Energy} = E_{\text{total}}^{(\text{DFT})} + \sum_{\substack{\text{2nd-row}\\ \text{atoms}}} \Delta E_{\text{x}}^{(\text{core})}. \tag{20}$$

For example, $Si_2H_6$ contains 2 silicon atoms. Its corrected total energy is $E_{\text{total}}^{(\text{DFT})} + 2 \times$ $(-21.1\ \text{kcal/mol})$.

The fact that this atom-wise correction scheme works very well implies that the errors stem from the core electrons which in general do not participate in chemical bonding. If so, the DFT energies without any correction should work well when predicting the bond energies, which is the case as shown in Fig. 4.

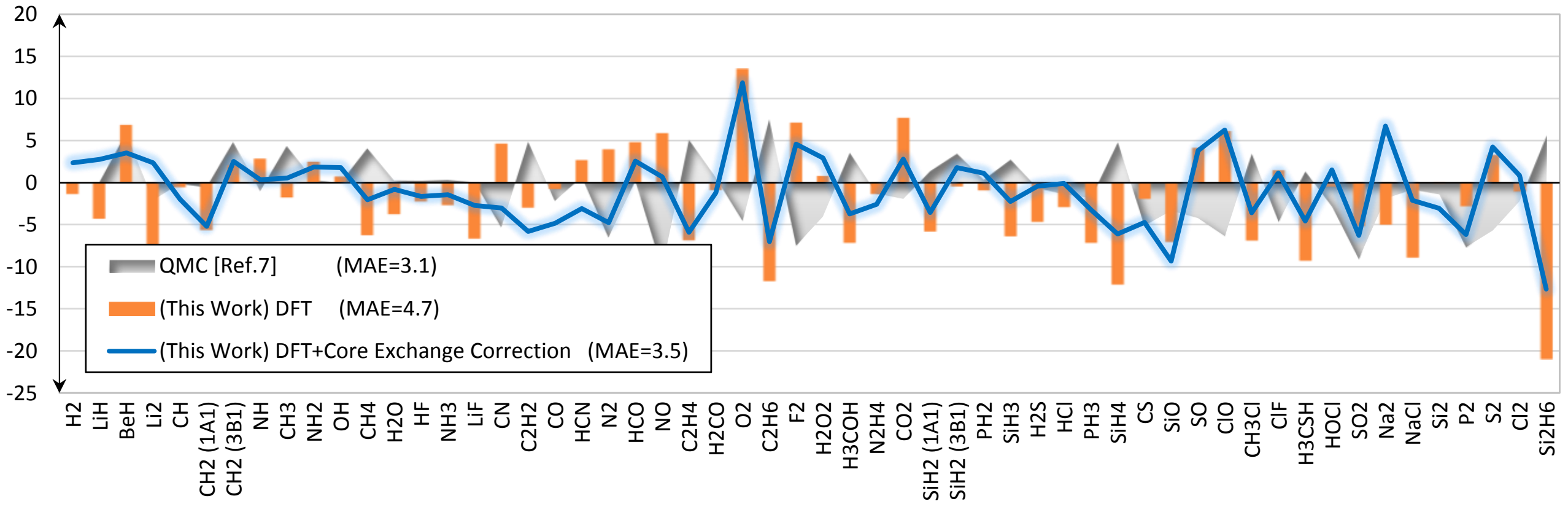

FIG 4. Errors of calculated bond energies (kcal/mol). The reference values are from Ref. 12.

Bond energies are the atomization energies without zero point energies nor relativistic contributions. We report the bond energies to allow direct comparisons with the reported QMC results.[7]

$$\text{Bond Energies} = \sum_{A \in \text{atoms}} \mathrm{E}_{\text{total}}^{(\text{A})} - E_{\text{total}}^{(\text{M})} \tag{21}$$

The summation runs through all constituent atoms in the molecule. In Fig. 4, under the label "(This Work) DFT", errors of the predicted bond energies are shown. The molecular total energies and the atomic total energies are from the DFT calculations without any correction. The MAE is 4.7 kcal/mol.

An alternative way to compute the bond energies is to use the "exact" atomic energies, as supposed to the same method that has been used to compute the molecular energy. Typically, this will lead to huge errors if we cannot predict the molecular total energies accurately enough. It is interesting to see how the corrected molecular energies perform in this test. The results are shown in Fig. 4 under the label "(This Work) DFT+Core Exchange Correction". In this case, the molecular total energies that enter Eq. (21) are the corrected ones; whereas the atomic total energies are the exact values from Ref. 12. The MAE now drops to 3.5 kcal/mol, very close to that of the QMC results. In fact, the 3.5 kcal/mol MAE for the bond energies is identical to that of the total energies. This is no coincidence as one can mathematically show that both errors are identical, provided that we use the "exact" atomic energies to evaluate the bond energies. The proof is provided in the supplementary material.

TABLE I. Errors of the calculated total energies, bond energies, dipole moments (Debye), and zero point energies of test molecules. All energies are in kcal/mol. Experimental dipole moments and zero point energies are from Ref. 17.

| | Eq. (17-18) | Eq. (17-18) $+\Delta E_x^{(\mathrm{core})}$ |
|---|---|---|
| *Total Energy* | | |
| ME | 16.9 | 1.0 |
| MAE | 18.5 | 3.5 |
| first-row ME | 0.4 | 0.4 |
| first-row MAE | 3.2 | 3.2 |
| *Bond Energy* | | |
| ME | -1.9 | -1.0 |
| MAE | 4.7 | 3.5 |
| first-row ME | -0.3 | -0.4 |
| first-row MAE | 4.4 | 3.2 |
| *Dipole Moment* | | |
| MAE | 0.11 | |
| *Zero Point Energy* | | |
| MAE (6-31G*) | 0.12 | |

In addition to the energies previously discussed, one may look at other physical properties such as dipole moments or zero point energies. Table I shows the errors of calculated dipole moments which reflect the accuracy of the predicted electron density (i.e. the x-component of a dipole moment involves an integral $\int x\rho(\vec{r})d^3r$). The MAE is 0.11 Debye which is comparable to that of other DFT functionals.[18]

We believe zero point energies (ZPE) are indicators of how well the functional predicts the derivatives of energies. The first derivatives are needed for geometry optimization. At the optimized geometries, the second derivatives are needed to compute vibration frequencies; and the sum of those frequencies are then translated to the vibrational ZPEs. Due to our limited computational resources, we can only use the small 6-31G* basis set. From Table I, the MAE is 0.12 kcal/mol which is comparable to that of the highly cited technique[19] of predicting ZPEs. Therefore, in addition to the total energies, the functionals in Eq. (17) and Eq. (18) are able to predict the derivatives of energies quite accurately.

The accuracies of Eq. (17) and Eq. (18) are perplexing when considering how much total energies are involved in the calculations. Despite nearly 1,000 folds increase in total energies, 739 kcal/mol for $H_2$ to more than half a million kcal/mol for $Cl_2$, the errors in Fig. 3 remain flat on the order of a few kcal/mol without apparent over-shooting nor under-shooting. From this data set, it is reasonable to project that even if the molecules are much larger, the errors should remain roughly the same. In other words, the DFT exchange-correlation functinoal combo has no systematic error. Even if the molecules contain second-row atoms, the energies can still be corrected very accurately.

For molecules, the most accurate all-electron total energy predictions are from QMC method such as the one by Nemec and coworkers,[7] where a set of 55 molecules were studied. Using the data from their supplementary material, we computed the MAE to be 13.5 kcal/mol for the first-row molecules, and 23.7 kcal/mol for the full set. A slightly more accurate (much smaller set of molecules) QMC total energies were also reported.[20] The MAE for the first-row homonuclear diatomic molecules was 11.4 kcal/mol. As evident from Fig. 3 and Table I, the DFT results in this work are the most accurate molecular total energy predictions to date, representing significant advances in the field of quantum mechanics.

The success of Eq. (17) and Eq. (18) does not imply that these functionals perform better than QMC method in general. QMC method has been shown to produce very accurate atomic energies with only 1.5 kcal/mol MAE;[21] but the same level of accuracy has not been achieved for molecules. QMC calculations generally require huge computing power. For example, the QMC energies shown in Fig. 3 required 40 CPU hours per electron (Intel XEON, 3 GHz CPU).[7] In contrast, our DFT calculations used 10 CPU hours to complete the full set of 56 molecules, translating to 0.01 CPU hours per electron (Intel Celeron, 1.86 GHz). Therefore, if one needs a speedy result with an overall error of a few kcal/mol, this functional combo provides a good alternative to the existing theoretical methods.

While preparing the manuscript, the exchange functional has drawn interest from a few authors. It was implemented in the LibXC[16] under the name “Chachiyo exchange”. It was also used[22] to compute atomic potentials for a new initial guess method called SAP (Superposition of Atomic

Potential). The study used 259 test molecules and reported that the Chachiyo exchange yielded the best initial guess wave functions on average out of hundreds functionals available in LibXC.

## IV. CONCLUSIONS

In this study, we show that a simple and accurate exchange functional can be devised using a model that a free-electron gas is perturbed by the periodic external potential which also induces Bragg diffraction of the Fermi electrons. Using this model, we have determined the quadratic coefficient of the exchange enhancement factor to be $\mu = \frac{8}{27}$, which governs how the exchange energy is enhanced by electron density gradient in the slowly varying limit. We also show that the slowly varying limit can be merged into the asymptotic limit using an interpolation scheme, yielding the complete exchange energy functional applicable to the full range of electron density gradients. The interpolating function is $w(s) = \frac{1}{ds+1}$, which mimics the exact exchange functional of a hydrogen atom.

Tests on 56 small molecules indicate that the exchange energy functional, in combination with the recently published correlation functional, is uniquely accurate when predicting the total energies of molecules, with the MAE of only 3.2 kcal/mol for the first-row molecules. For the second-row molecules, errors of the core electron exchange energies can be corrected, leading to the most accurate molecular total energy predictions to date despite minimal computational efforts.

For bond energies, dipole moments, and zero point energies, the MAEs are 3.5 kcal/mol, 0.11 Debye, and 0.12 kcal/mol, respectively. Eq. (17) and Eq. (18) are able to capture the behavior of valent electrons quite well as seen from the successful bond energy calculations. Accurate dipole moments reflect how well the functionals predict the electron density on average; whereas the zero point energies indicate the performance of first and second derivatives of energy predictions.

In DFT, the basic variable is the density of electron, a function that holds very little mathematical information compared to the full multi-electron wave functions, yet theoretically capable of determining exactly the ground-state total energies of electronic systems. The exchange energy functional in Eq. (17) is a practical demonstration that the basic premise of DFT is valid. Its non-empirical form, and its unique accuracy when predicting the total energies of molecules represent a mile stone in quantum mechanics, in the pursuit to understand how the energies could be determined exactly by the density of electron.

## Supplementary Material

See supplementary material for additional theoretical proofs, detailed energies, dipole moments, and zero point energies of the test molecules.

## Acknowledgements

Support by Thailand Center of Excellence in Physics grant No. ThEP-60-PET-NU9 is gratefully acknowledged.

TABLE I. Errors of the calculated total energies, bond energies, dipole moments (Debye), and zero point energies of test molecules. All energies are in kcal/mol. Experimental dipole moments and zero point energies are from Ref. 17.

| | Eq. (17-18) | Eq. (17-18) $+\Delta E_x^{(\mathrm{core})}$ |
|---|---|---|
| *Total Energy* | | |
| ME | 16.9 | 1.0 |
| MAE | 18.5 | 3.5 |
| first-row ME | 0.4 | 0.4 |
| first-row MAE | 3.2 | 3.2 |
| *Bond Energy* | | |
| ME | -1.9 | -1.0 |
| MAE | 4.7 | 3.5 |
| first-row ME | -0.3 | -0.4 |
| first-row MAE | 4.4 | 3.2 |
| *Dipole Moment* | | |
| MAE | 0.11 | |
| *Zero Point Energy* | | |
| MAE (6-31G*) | 0.12 | |

# Supplementary Material

## Simple and accurate exchange energy for density functional theory

Teepanis Chachiyo,[1,2,*] and Hathaithip Chachiyo[3]

[1] Department of Physics, Faculty of Science, Naresuan University, Phitsanulok 65000 Thailand
email: teepanisc@nu.ac.th

[2] Thailand Center of Excellence in Physics, Ministry of Higher Education, Science, Research and Innovation, 328 Si Ayutthaya Road, Bangkok 10400, Thailand

[3] Private Residence, Phitsanulok 65000, Thailand        email: hathaithip.chachiyo@gmail.com

[*] Corresponding author

**Acknowledgements**

Support by Thailand Center of Excellence in Physics grant No. ThEP-60-PET-NU9 is gratefully acknowledged.

Page

## Proof that $\frac{2K^2[\rho(\vec{K})]^2}{|\vec{\nabla}\rho|^2} = 1$ on average

Upon perturbation by a periodic external potential $V_{\text{ext}}(\vec{K})e^{+i\vec{K}\cdot\vec{r}} + V_{\text{ext}}(-\vec{K})e^{-i\vec{K}\cdot\vec{r}}$ , the free-electron gas density becomes

$$\rho = \rho_0 + \rho(\vec{K})e^{+i\vec{K}\cdot\vec{r}} + \rho(-\vec{K})e^{-i\vec{K}\cdot\vec{r}} = \rho_0 + 2\rho(\vec{K})\cos(\vec{K}\cdot\vec{r}), \tag{S1}$$

which leads to the gradient of electron density

$$\vec{\nabla}\rho = -2\rho(\vec{K})\sin(\vec{K}\cdot\vec{r})\vec{K} \tag{S2}$$

and

$$|\vec{\nabla}\rho|^2 = 4K^2[\rho(\vec{K})]^2\sin^2(\vec{K}\cdot\vec{r}). \tag{S3}$$

Therefore, the average of $|\vec{\nabla}\rho|^2$ over the normalization volume $\Omega$ is

$$\frac{1}{\Omega}\int|\vec{\nabla}\rho|^2 d^3r = \frac{1}{\Omega}4[\rho(\vec{K})]^2 K^2\int\sin^2(\vec{K}\cdot\vec{r})d^3r. \tag{S4}$$

Since the volume is periodic, replacing the integrand with $\cos^2(\vec{K}\cdot\vec{r})$ would only shift the phase, but the integrated value would not change. In other words,

$$\int\sin^2(\vec{K}\cdot\vec{r})d^3r = \int\cos^2(\vec{K}\cdot\vec{r})d^3r = \frac{1}{2}\int\left[\cos^2(\vec{K}\cdot\vec{r}) + \sin^2(\vec{K}\cdot\vec{r})\right]d^3r = \frac{\Omega}{2}. \tag{S5}$$

Hence,

$$\frac{1}{\Omega}\int\left|\vec{\nabla}\rho\right|^2 d^3r = 2K^2\left[\rho\left(\vec{K}\right)\right]^2. \tag{S6}$$

Consider $\frac{2K^2\left[\rho\left(\vec{K}\right)\right]^2}{\left|\vec{\nabla}\rho\right|^2}$. If we replace the term $\left|\vec{\nabla}\rho\right|^2$ by its average value as shown in the above expression, we have

$$\frac{2K^2\left[\rho\left(\vec{K}\right)\right]^2}{\left|\vec{\nabla}\rho\right|^2} = 1 \textit{ on average}. \tag{S7}$$

**Proof that the errors of total energies and bond energies are the same (magnitude)**

Let $E_{\text{exact}}^{(\text{M})}$ and $E_{\text{exact}}^{(\text{A})}$ be the exact total energies of molecules and of atoms, respectively. If a theory predicts the total energy of a molecule to be $E_{\text{theory}}^{(\text{M})}$ , then the error is equal to

$$\Delta E^{(\text{M})} = E_{\text{theory}}^{(\text{M})} - E_{\text{exact}}^{(\text{M})}. \tag{S8}$$

A (positive value) bond energy is computed by subtracting the molecular total energy from the sum of atomic energies.

$$E_{\text{exact}}^{(\text{Bond})} = \sum_{\text{A}\in\text{atoms}} E_{\text{exact}}^{(\text{A})} - E_{\text{exact}}^{(\text{M})}. \tag{S9}$$

For a theoretical prediction of bond energy, both the molecular energy and the atomic energies are from the same theory, relying on a certain degree of error cancellation, even though the $E_{\text{theory}}^{(\text{M})}$ and $E_{\text{theory}}^{(\text{A})}$ are both off by huge amount, the difference is still quite accurate.

However, let us evaluate the bond energy using

$$E_{\text{theory}}^{(\text{Bond})} = \sum_{\text{A}\in\text{atoms}} E_{\text{exact}}^{(\text{A})} - E_{\text{theory}}^{(\text{M})}. \tag{S10}$$

Notice how we are using the exact atomic energies instead of the theoretical ones. The error of the bond energy in Eq. (S10) is

$$\begin{aligned} \Delta E^{(\text{Bond})} &= E_{\text{theory}}^{(\text{Bond})} - E_{\text{exact}}^{(\text{Bond})} \\ &= \left(\sum_{\text{A}\in\text{atoms}} E_{\text{exact}}^{(\text{A})} - E_{\text{theory}}^{(\text{M})}\right) - E_{\text{exact}}^{(\text{Bond})}. \end{aligned} \tag{S11}$$

Substituting $E_{\text{exact}}^{(\text{Bond})} = \sum_{\text{A}\in\text{atoms}} E_{\text{exact}}^{(\text{A})} - E_{\text{exact}}^{(\text{M})}$ from Eq. (S9) into the above expression yields,

$$\Delta E^{(\text{Bond})} = \left(\sum_{\text{A}\in\text{atoms}} E_{\text{exact}}^{(\text{A})} - E_{\text{theory}}^{(\text{M})}\right) - \left(\sum_{\text{A}\in\text{atoms}} E_{\text{exact}}^{(\text{A})} - E_{\text{exact}}^{(\text{M})}\right) \tag{S12}$$

The atomics energies cancel and we are left with

$$\Delta E^{(\text{Bond})} = -E_{\text{theory}}^{(\text{M})} + E_{\text{exact}}^{(\text{M})}, \tag{S13}$$

which is precisely the negative of Eq. (S8). Therefore,

$$\Delta E^{(\text{M})} = -\Delta E^{(\text{Bond})} \tag{S14}$$

In other words, the error of predicted the molecular total energies is of the same magnitude as that of the bond energies prediction, <u>provided that we use the exact atomic energies as in Eq. (S10).</u>

# Calculated Energies

Reference total energies used to compute the errors are from Ref. [1].

## 1) Atoms

**Color Code**
**Calculated Values** << Copied by values from Siam Quantum
**Analysis** << Worksheet formula

| name | (This Work) Etotal QZP-gh [Ha] | ---- Core Exchange Correction Scheme for Ne and 2nd-row atoms ----- | | | | | | |
|---|---|---|---|---|---|---|---|---|
| | | | Exchange Energy @HF Density | | | (This Work) | | |
| | | Core Config. | DFT ex. | HF ex. | Correction [Ha] | Corected Energies [Ha] | | |
| H | -0.502981 | | | | 0 | -0.502981 | (No Correction) | |
| He | -2.908144 | | | | 0 | -2.908144 | (No Correction) | |
| Li | -7.486352 | | | | 0 | -7.486352 | (No Correction) | |
| Be | -14.659224 | | | | 0 | -14.659224 | (No Correction) | |
| B | -24.647743 | | | | 0 | -24.647743 | (No Correction) | |
| C | -37.839799 | | | | 0 | -37.839799 | (No Correction) | |
| N | -54.582274 | | | | 0 | -54.582274 | (No Correction) | |
| O | -75.066050 | | | | 0 | -75.066050 | (No Correction) | |
| F | -99.731913 | | | | 0 | -99.731913 | (No Correction) | |
| Ne | -128.928050 | Ne8+ [1s] | -6.0070 | -6.0277 | -0.0207 | -128.948712 | | |
| Na | -162.240410 | Na9+ [1s] | -6.6291 | -6.6526 | -0.0235 | -162.263942 | | |
| Mg | -200.026734 | Mg10+[1s] | -7.2511 | -7.2776 | -0.0265 | -200.053214 | | |
| Al | -242.317156 | Al11+[1s] | -7.8733 | -7.9027 | -0.0294 | -242.346556 | | |
| Si | -289.322973 | Si4+[1s2s2p] | -19.1780 | -19.2117 | -0.0337 | -289.356653 | | |
| P | -341.212425 | P5+[1s2s2p] | -20.9280 | -20.9719 | -0.0439 | -341.256335 | | |
| S | -398.057341 | S6+[1s2s2p] | -22.6763 | -22.7297 | -0.0535 | -398.110801 | | |
| Cl | -460.087376 | Cl7+[1s2s2p] | -24.4236 | -24.4858 | -0.0621 | -460.149526 | | |
| Ar | -527.466732 | Ar8+[1s2s2p] | -26.1695 | -26.2403 | -0.0709 | -527.537593 | | |

2) 56 Molecules.

| name | (This Work) Etotal QZP-gh [Ha] | (This Work) Corected Etot QZP-gh [Ha] | (This Work) Ebond QZP-gh [kcal/mol] | (This Work) Corrected Ebond QZP-gh [kcal/mol] |
|---|---|---|---|---|
| H2 | -1.178148 | -1.178148 | 108.04833 | 111.789543 |
| LiH | -8.074777 | -8.074777 | 53.616912 | 60.66572539 |
| BeH | -15.252419 | -15.252419 | 56.610132 | 53.35022091 |
| Li2 | -14.998953 | -14.998953 | 16.471494 | 26.82790899 |
| CH | -38.475660 | -38.475660 | 83.383448 | 81.99037703 |
| CH2 (1A1) | -39.126332 | -39.126332 | 176.06094 | 176.538472 |
| CH2 (3B1) | -39.152413 | -39.152413 | 192.42701 | 192.9045444 |
| NH | -55.223226 | -55.223226 | 86.578098 | 84.10257364 |
| CH3 | -39.836455 | -39.836455 | 306.04449 | 308.3926278 |
| NH2 | -55.882231 | -55.882231 | 184.48462 | 183.8797044 |
| OH | -75.739939 | -75.739939 | 107.24637 | 108.3325938 |
| CH4 | -40.512545 | -40.512545 | 414.67201 | 418.8907564 |
| H2O | -76.437104 | -76.437104 | 229.09866 | 232.0554828 |
| HF | -100.456684 | -100.456684 | 139.17531 | 139.7990522 |
| NH3 | -56.562424 | -56.562424 | 295.68682 | 296.952504 |
| LiF | -107.430160 | -107.430160 | 132.9661 | 136.8974487 |
| CN | -92.720175 | -92.720175 | 187.0618 | 179.4519981 |
| C2H2 | -77.326369 | -77.326369 | 402.11367 | 399.3275236 |
| CO | -113.318379 | -113.318379 | 258.86645 | 254.818386 |
| HCN | -93.426175 | -93.426175 | 314.45813 | 308.7189326 |
| N2 | -109.534502 | -109.534502 | 232.14961 | 223.4573492 |
| HCO | -113.861188 | -113.861188 | 283.85889 | 281.6814355 |
| NO | -129.905881 | -129.905881 | 161.61944 | 156.4889194 |
| C2H4 | -78.579291 | -78.579291 | 557.08338 | 558.0384538 |
| H2CO | -114.507057 | -114.507057 | 373.52245 | 373.2156029 |
| O2 | -150.345597 | -150.345597 | 133.97137 | 132.402599 |
| C2H6 | -79.816232 | -79.816232 | 702.02488 | 706.7211564 |
| F2 | -199.537481 | -199.537481 | 46.219204 | 43.72548187 |
| H2O2 | -151.568159 | -151.568159 | 269.88991 | 272.0623441 |
| H3COH | -115.724642 | -115.724642 | 506.31787 | 509.7522337 |
| N2H4 | -111.873459 | -111.873459 | 437.36589 | 436.1560498 |
| CO2 | -188.605665 | -188.605665 | 397.69412 | 392.8616669 |
| SiH2 (1A1) | -290.562529 | -290.596209 | 146.58243 | 148.8508751 |
| SiH2 (3B1) | -290.538366 | -290.572046 | 131.41992 | 133.6883657 |
| PH2 | -342.461371 | -342.505281 | 152.47474 | 154.5436403 |
| SiH3 | -291.184418 | -291.218098 | 221.19832 | 225.3373673 |
| H2S | -399.348422 | -399.401882 | 178.91485 | 183.1586961 |
| HCl | -460.756891 | -460.819041 | 104.50165 | 107.3298337 |
| PH3 | -343.097163 | -343.141073 | 235.81489 | 239.7543955 |
| SiH4 | -291.831002 | -291.864682 | 311.31055 | 317.3202039 |
| CS | -436.168105 | -436.221565 | 170.03308 | 167.2720408 |
| SiO | -364.685259 | -364.718939 | 185.89087 | 183.6337207 |
| SO | -473.330888 | -473.384348 | 130.20632 | 129.9245644 |
| ClO | -535.267984 | -535.330134 | 71.886221 | 72.05941341 |
| CH3Cl | -500.056311 | -500.118461 | 389.17693 | 392.4826512 |
| ClF | -559.921683 | -559.983833 | 64.253197 | 63.96391475 |
| H3CSH | -438.651663 | -438.705123 | 465.98785 | 470.7092269 |
| HOCl | -535.919752 | -535.981902 | 165.25146 | 167.2952587 |
| SO2 | -548.595256 | -548.648716 | 254.65272 | 253.5865851 |
| Na2 | -324.499718 | -324.546782 | 11.858672 | 23.58305793 |
| NaCl | -622.471571 | -622.557253 | 90.226438 | 97.04620985 |
| Si2 | -578.766347 | -578.833707 | 75.552758 | 72.6072291 |
| P2 | -682.606650 | -682.694470 | 114.08121 | 110.7365822 |
| S2 | -796.284027 | -796.390947 | 106.26558 | 107.2708479 |
| Cl2 | -920.267031 | -920.391331 | 57.905939 | 59.82109775 |
| Si2H6 | -582.479772 | -582.547132 | 512.01001 | 520.2881164 |

# Errors of the Calculated Energies

Refer to the supplemental material of Ref. [2] for the reported QMC total energies and bond energies. We calculated the errors as compared to the experimentally derived energies from Ref. [1].

1) Atoms

| | Error of Etot | | |
|---|---|---|---|
| | As IS | Corrected | DMC |
| | QZP-gh [kcal/mol] | QZP-gh [kcal/mol] | Nemec2010 [kcal/mol] |
| name | | | |
| H | -1.9 | -1.9 | 0.0 |
| He | -2.8 | -2.8 | 0.0 |
| Li | -5.2 | -5.2 | 0.1 |
| Be | 5.1 | 5.1 | 6.3 |
| B | 3.9 | 3.9 | 8.6 |
| C | 3.3 | 3.3 | 9.3 |
| N | 4.3 | 4.3 | 7.9 |
| O | 0.8 | 0.8 | 9.3 |
| F | 1.2 | 1.2 | 9.9 |
| Ne | 6.0 | -7.0 | 8.9 |
| Na | 8.9 | -5.9 | 9.4 |
| Mg | 16.5 | -0.1 | 13.1 |
| Al | 18.1 | -0.3 | 15.8 |
| Si | 22.6 | 1.5 | 19.7 |
| P | 29.2 | 1.7 | 22.9 |
| S | 33.0 | -0.5 | 27.7 |
| Cl | 38.0 | -1.0 | 33.3 |
| Ar | 46.0 | 1.5 | 36.8 |
| **Average** | 12.6 | -0.1 | 13.3 |
| **MAE** | 13.7 | 2.7 | 13.3 |
| **First row MAE** | 3.4 | 3.5 | 6.0 |
| **First row Avg** | 1.5 | 0.2 | 6.0 |

## 2) Molecules

| | Error of Etot | | | Error of E_bond | | |
|---|---|---|---|---|---|---|
| | As IS | Corrected | DMC | As IS | Corrected | DMC |
| | **QZP-gh [kcal/mol]** | **QZP-gh [kcal/mol]** | **Nemec2010 [kcal/mol]** | **QZP-gh [kcal/mol]** | **QZP-gh [kcal/mol]** | **Nemec2010 [kcal/mol]** |
| **name** | | | | | | |
| H2 | -2.3 | -2.3 | | -1.5 | 2.3 | |
| LiH | -2.7 | -2.7 | 0.0 | -4.3 | 2.7 | 0.0 |
| BeH | -3.5 | -3.5 | 0.5 | 6.8 | 3.5 | 5.8 |
| Li2 | -2.4 | -2.4 | 2.1 | -7.9 | 2.4 | -2.0 |
| CH | 2.0 | 2.0 | 9.4 | -0.6 | -2.0 | -0.1 |
| CH2 (1A1) | 5.2 | 5.2 | 9.8 | -5.7 | -5.2 | -0.5 |
| CH2 (3B1) | -2.5 | -2.5 | 4.5 | 2.0 | 2.5 | 4.8 |
| NH | -0.3 | -0.3 | 8.9 | 2.8 | 0.3 | -1.0 |
| CH3 | -0.6 | -0.6 | 4.9 | -1.7 | 0.6 | 4.4 |
| NH2 | -1.8 | -1.8 | 7.7 | 2.4 | 1.8 | 0.2 |
| OH | -1.8 | -1.8 | 9.3 | 0.7 | 1.8 | 0.0 |
| CH4 | 2.0 | 2.0 | 5.2 | -6.3 | -2.0 | 4.1 |
| H2O | 0.8 | 0.8 | 9.0 | -3.7 | -0.8 | 0.3 |
| HF | 1.6 | 1.6 | 9.6 | -2.2 | -1.6 | 0.3 |
| NH3 | 1.4 | 1.4 | 7.6 | -2.7 | -1.4 | 0.3 |
| LiF | 2.7 | 2.7 | 9.8 | -6.6 | -2.7 | 0.1 |
| CN | 3.0 | 3.0 | 22.5 | 4.6 | -3.0 | -5.3 |
| C2H2 | 5.7 | 5.7 | 13.7 | -2.9 | -5.7 | 4.9 |
| CO | 4.8 | 4.8 | 20.8 | -0.8 | -4.8 | -2.2 |
| HCN | 3.1 | 3.1 | 16.5 | 2.6 | -3.1 | 0.7 |
| N2 | 4.8 | 4.8 | 22.4 | 3.9 | -4.8 | -6.5 |
| HCO | -2.6 | -2.6 | 18.3 | 4.8 | 2.6 | 0.3 |
| NO | -0.7 | -0.7 | 27.3 | 5.9 | 0.7 | -10.1 |
| C2H4 | 6.0 | 6.0 | 13.5 | -6.9 | -6.0 | 5.0 |
| H2CO | 1.2 | 1.2 | 18.0 | -0.9 | -1.2 | 0.6 |
| O2 | -11.9 | -11.9 | 23.1 | 13.4 | 11.9 | -4.5 |
| C2H6 | 7.0 | 7.0 | 11.1 | -11.7 | -7.0 | 7.5 |
| F2 | -4.5 | -4.5 | 27.3 | 7.0 | 4.5 | -7.5 |
| H2O2 | -2.9 | -2.9 | 22.6 | 0.8 | 2.9 | -4.0 |
| H3COH | 3.7 | 3.7 | 15.0 | -7.2 | -3.7 | 3.6 |
| N2H4 | 2.6 | 2.6 | 17.2 | -1.4 | -2.6 | -1.3 |
| CO2 | -2.9 | -2.9 | 29.7 | 7.7 | 2.9 | -1.9 |
| SiH2 (1A1) | 24.7 | 3.6 | 18.4 | -5.8 | -3.6 | 1.4 |
| SiH2 (3B1) | 19.3 | -1.8 | 16.3 | -0.5 | 1.8 | 3.4 |
| PH2 | 26.4 | -1.1 | 22.5 | -1.0 | 1.1 | 0.4 |
| SiH3 | 23.4 | 2.3 | 17.0 | -6.4 | -2.3 | 2.7 |
| H2S | 34.0 | 0.5 | 28.3 | -4.7 | -0.5 | -0.6 |
| HCl | 39.1 | 0.1 | 34.7 | -2.9 | -0.1 | -1.4 |
| PH3 | 30.8 | 3.2 | 24.7 | -7.2 | -3.2 | -1.8 |
| SiH4 | 27.2 | 6.1 | 14.9 | -12.1 | -6.1 | 4.8 |
| CS | 38.3 | 4.7 | 42.1 | -2.0 | -4.7 | -5.1 |
| SiO | 30.4 | 9.3 | 32.3 | -7.0 | -9.3 | -3.3 |
| SO | 29.8 | -3.8 | 41.2 | 4.1 | 3.8 | -4.2 |
| ClO | 32.7 | -6.3 | 48.9 | 6.1 | 6.3 | -6.3 |
| CH3Cl | 42.7 | 3.7 | 39.2 | -7.0 | -3.7 | 3.4 |
| ClF | 37.8 | -1.2 | 47.9 | 1.4 | 1.2 | -4.6 |
| H3CSH | 38.1 | 4.6 | 35.7 | -9.3 | -4.6 | 1.3 |
| HOCl | 37.5 | -1.5 | 45.5 | -0.5 | 1.5 | -2.9 |
| SO2 | 39.9 | 6.3 | 55.5 | -5.3 | -6.3 | -9.1 |
| Na2 | 22.8 | -6.8 | 20.6 | -5.0 | 6.8 | -1.8 |
| NaCl | 55.8 | 2.0 | 43.4 | -8.9 | -2.0 | -0.7 |
| Si2 | 45.3 | 3.1 | 40.8 | -0.1 | -3.1 | -1.4 |
| P2 | 61.3 | 6.2 | 53.5 | -2.8 | -6.2 | -7.7 |
| S2 | 62.9 | -4.2 | 61.1 | 3.2 | 4.2 | -5.7 |
| Cl2 | 77.1 | -0.9 | 68.8 | -1.0 | 0.9 | -2.2 |
| Si2H6 | 54.9 | 12.6 | 33.7 | -20.9 | -12.6 | 5.7 |
| **Average** | 16.9 | 1.0 | 23.7 | -1.9 | -1.0 | -0.7 |
| **MAE** | 18.5 | 3.5 | 23.7 | 4.7 | 3.5 | 3.1 |
| **first-row AVG** | 0.4 | 0.4 | 13.5 | -0.3 | -0.4 | -0.1 |
| **first-row MAE** | 3.2 | 3.2 | 13.5 | 4.4 | 3.2 | 2.9 |
| **second-row AVG** | 38.8 | 1.7 | 37.0 | -4.0 | -1.7 | -1.5 |
| **second-row MAE** | 38.8 | 4.0 | 37.0 | 5.2 | 4.0 | 3.4 |

3) Other QMC results. Refer to Ref. [3] and Ref. [4] for the reported QMC total energies. Here we computed the errors in kcal/mol as compared to the exact energies (quoted in the references).

| name | QMC Type | | [Toulouse2008] QMC - exact [kcal/mol] |
|---|---|---|---|
| Li2 | DMC J×FVCAS | | 0.3 |
| Be2 | DMC J×FVCAS | | 0.4 |
| B2 | DMC J×FVCAS | | 5.2 |
| C2 | DMC J×FVCAS | | 10.0 |
| N2 | DMC J×FVCAS | | 13.9 |
| O2 | DMC J×FVCAS | | 20.7 |
| F2 | DMC J×FVCAS | | 21.0 |
| Ne2 | DMC J×SD | | 17.7 |
| | | **MAE** | 11.14 |

| | QMC Type | | [Brown2007] QMC - exact [kcal/mol] |
|---|---|---|---|
| Li | DMC2-MDBF | | 0.00 |
| Be | DMC2-MDBF | | 0.03 |
| B | DMC2-MDBF | | 0.21 |
| C | DMC2-MDBF | | 0.72 |
| N | DMC2-MDBF | | 1.19 |
| O | DMC2-MDBF | | 3.51 |
| F | DMC2-MDBF | | 4.08 |
| Ne | DMC2-MDBF | | 2.32 |
| | | **MAE** | 1.51 |

# Dipole Moments and Zero Point Energies

| Dipole | | | ZPE | | | |
|---|---|---|---|---|---|---|
| (This Work) QZP-gh Calc. [Debye] | | (This Work) QZP-gh Calc. - exp [Debye] | exp. ZPE [kcal/mol] | ----------- This Work --------- 6-31G* ZPE [Ha] | 6-31G* ZPE [kcal/mol] | 6-31G* Calc. - exp [kcal/mol] |
| 0.0000 | | | 6.2 | 0.00996 | 6.2 | 0.0 |
| 5.7155 | | -0.1685 | 2.0 | 0.00307 | 1.9 | -0.1 |
| 0.2744 | | | 2.9 | 0.00455 | 2.9 | -0.1 |
| 0.0000 | | | 0.5 | 0.00073 | 0.5 | 0.0 |
| 1.4316 | | -0.0284 | 4.0 | 0.00614 | 3.9 | -0.2 |
| 1.7713 | | | | 0.01600 | 10.0 | |
| 0.5995 | | | | 0.01705 | 10.7 | |
| 1.4964 | | 0.1064 | 4.6 | 0.00716 | 4.5 | -0.1 |
| 0.0000 | | | | 0.02916 | 18.3 | |
| 1.7602 | | | 11.5 | 0.01844 | 11.6 | 0.1 |
| 1.6224 | | -0.0326 | 5.3 | 0.00807 | 5.1 | -0.2 |
| 0.0000 | | | 27.1 | 0.04417 | 27.7 | 0.6 |
| 1.8436 | | -0.0110 | 12.9 | 0.02065 | 13.0 | 0.1 |
| 1.7639 | | -0.0623 | 5.9 | 0.00884 | 5.5 | -0.3 |
| 1.5085 | | 0.0367 | 20.6 | 0.03373 | 21.2 | 0.5 |
| 6.0996 | | -0.2278 | 1.3 | 0.00218 | 1.4 | 0.1 |
| 1.1573 | | | 2.9 | 0.00475 | 3.0 | 0.0 |
| 0.0002 | | | 15.3 | 0.02459 | 15.4 | 0.1 |
| 0.1849 | | 0.0751 | 3.1 | 0.00481 | 3.0 | -0.1 |
| 2.9238 | | -0.0614 | 8.7 | 0.01422 | 8.9 | 0.2 |
| 0.0006 | | | 3.4 | 0.00535 | 3.4 | 0.0 |
| 1.3894 | | | 7.8 | 0.01259 | 7.9 | 0.1 |
| 0.2064 | | 0.0477 | 2.7 | 0.00431 | 2.7 | 0.0 |
| 0.0000 | | | | 0.04981 | 31.3 | |
| 2.2030 | | -0.1290 | 16.1 | 0.02588 | 16.2 | 0.1 |
| 0.0002 | | | 2.3 | 0.00356 | 2.2 | 0.0 |
| 0.0000 | | | | 0.07341 | 46.1 | |
| 0.0000 | | | 1.3 | 0.00228 | 1.4 | 0.1 |
| 1.7142 | | 0.1412 | | 0.02535 | 15.9 | |
| 1.5683 | | -0.1317 | | 0.05006 | 31.4 | |
| 1.8773 | | 0.1273 | | 0.05197 | 32.6 | |
| 0.0002 | | | 6.2 | 0.00979 | 6.1 | -0.1 |
| 0.2229 | | | 7.2 | 0.01132 | 7.1 | -0.1 |
| 0.0490 | | | | 0.01180 | 7.4 | |
| 0.6047 | | | | 0.01310 | 8.2 | |
| 0.0239 | | | | 0.02069 | 13.0 | |
| 1.0313 | | 0.0530 | 9.2 | 0.01479 | 9.3 | 0.1 |
| 1.1216 | | 0.0130 | 4.2 | 0.00657 | 4.1 | -0.1 |
| 0.6647 | | 0.0907 | 14.6 | 0.02354 | 14.8 | 0.2 |
| 0.0000 | | | 19.2 | 0.03035 | 19.0 | -0.2 |
| 1.9319 | | -0.0261 | 1.8 | 0.00282 | 1.8 | -0.1 |
| 2.8138 | | -0.2844 | 1.8 | 0.00269 | 1.7 | -0.1 |
| 1.3671 | | -0.1829 | 1.6 | 0.00243 | 1.5 | -0.1 |
| 1.3514 | | 0.0544 | 1.2 | 0.00187 | 1.2 | 0.0 |
| 1.8790 | | -0.0173 | | 0.03717 | 23.3 | |
| 0.7458 | | -0.1423 | 1.1 | 0.00170 | 1.1 | -0.1 |
| 1.5600 | | 0.0400 | | 0.04523 | 28.4 | |
| 1.5324 | | 0.2324 | 8.0 | 0.01263 | 7.9 | 0.0 |
| 1.5061 | | -0.1270 | 4.3 | 0.00639 | 4.0 | -0.3 |
| 0.0000 | | | 0.2 | 0.00034 | 0.2 | 0.0 |
| 8.6939 | | -0.3073 | 0.5 | 0.00078 | 0.5 | 0.0 |
| 0.0000 | | | 0.7 | 0.00121 | 0.8 | 0.0 |
| 0.0007 | | | 1.1 | 0.00175 | 1.1 | 0.0 |
| 0.0002 | | | 1.0 | 0.00154 | 1.0 | -0.1 |
| 0.0000 | | | 0.8 | 0.00115 | 0.7 | -0.1 |
| 0.0000 | | | | 0.04776 | 30.0 | |
| | **Avg E** | -0.03 | | | **Avg E** | -0.003 |
| | **MAE** | 0.11 | | | **MAE** | 0.120 |
| | **Count** | 28 | | | **Count** | 42 |

For dipole moments, refer to the experimental values from Ref. [5]. For the zero point energies, we calculated the "exp." values listed above from experimental vibration energies in Ref. [5]. For diatomic molecules where the anharmonic term $\omega_e x_e$ was available, we used

$$\mathrm{ZPE} = \frac{1}{2}\omega_e - \frac{1}{4}\omega_e x_e. \tag{S15}$$

For polyatomic molecules, we used

$$\mathrm{ZPE} = \sum_{\text{all modes}} \frac{1}{2}\omega_e. \tag{S16}$$

## Molecular Coordinates

For H2 we used experimental bond length of 0.741 angstroms. For other 55 molecules, the geometries could be downloaded from the supplementary material of Ref. [2], (doi: 10.1063/1.3288054)